# ExpertRAG: Efficient RAG with Mixture of Experts – Optimizing Context Retrieval for Adaptive LLM Responses

Ⓘ **Esmail Gumaan**\*

Faculty of Computer Science and Information Technology
University of Sana'a
Yemen, Sana'a
esmailG231601.ue.ye.edu

March 16, 2025

## ABSTRACT

*ExpertRAG* is a novel theoretical framework that integrates Mixture-of-Experts (MoE) architectures with Retrieval-Augmented Generation (RAG) to advance the efficiency and accuracy of knowledge-intensive language modeling. We propose a dynamic retrieval gating mechanism coupled with expert routing, enabling the model to selectively consult an external knowledge store or rely on specialized internal experts based on the query's needs. The paper lays out the theoretical foundations of ExpertRAG, including a probabilistic formulation that treats retrieval and expert selection as latent decisions, and mathematical justifications for its efficiency in both computation and knowledge utilization. We derive formulae to quantify the expected computational cost savings from selective retrieval and the capacity gains from sparse expert utilization. A comparative analysis positions ExpertRAG against standard RAG (with always-on retrieval) and pure MoE models (e.g. Switch Transformer, Mixtral) to highlight its unique balance between parametric knowledge and non-parametric retrieval. We also outline an experimental validation strategy, proposing benchmarks and evaluation protocols to test ExpertRAG's performance on factual recall, generalization, and inference efficiency. The proposed framework, although presented theoretically, is supported by insights from prior work in RAG and MoE, and is poised to provide more **factual**, **efficient**, and **adaptive** generation by leveraging the best of both paradigms. In summary, ExpertRAG contributes a new perspective on scaling and augmenting language models, backed by a thorough analysis and a roadmap for empirical validation.

## 1 Introduction

Large language models (LLMs) have achieved remarkable success in many NLP tasks, yet they face persistent challenges in knowledge-intensive applications. A key limitation is the reliance on storing factual knowledge purely in model parameters. As models grow, their ability to recall or update specific facts becomes problematic [1]. Retrieval-Augmented Generation (RAG) was introduced to address this by equipping models with access to external non-parametric memory (e.g. a text corpus or database), allowing them to fetch relevant information on-the-fly [2]. RAG combines a parametric neural generator with a retrieval module, producing outputs that are more specific and factual than those of parametric-only models [1]. However, standard RAG pipelines retrieve documents for every query, which can be inefficient when the model's internal knowledge is already sufficient. Unnecessary retrieval incurs extra latency and may introduce distractors, highlighting a need for dynamic retrieval strategies that invoke external lookup only when needed.

In parallel, Mixture-of-Experts (MoE) architectures have emerged as a solution for scaling model capacity without proportional increases in computation[3]. In an MoE model, multiple expert subnetworks (e.g. feed-forward layers) are trained, and a learned gating function routes each input token or example to a subset of these experts. This sparse activation means only a few experts process each token, allowing the total parameter count to increase (potentially to trillions) while keeping the compute per token comparable to a much smaller model [4]. Notable MoE instances include the Switch Transformer [4], which demonstrated that a 1.6-trillion-parameter model can be executed with the computational cost of a 1.4-billion-parameter dense model by routing each token to a single expert, and Mixtral (Mix-of-Mistral experts), an 8×7B MoE transformer that combines eight 7B expert models into a larger sparse model [5]. These architectures not only improve the efficiency–capacity trade-off but also enable a form of expert specialization, where different experts can learn different aspects of the data [3]. In principle, MoE models could handle diverse query types by routing to domain-specific experts, but like all parametric models, they remain limited to the knowledge present in their training data and parameters.

The motivation for integrating MoE with RAG stems from a desire to combine the strengths of external retrieval and expert specialization. We hypothesize that an architecture which can both route queries to appropriate internal experts and decide when to fetch external information would yield significant benefits: (1) Improved accuracy and factuality – The model can consult an external knowledge source for queries that fall outside its learned expertise, reducing hallucinations and errors on obscure facts while still leveraging parametric knowledge for in-domain questions [1]. (2) Efficiency via selective retrieval – By gating the retrieval mechanism, the model avoids the overhead of searching and reading documents when it is confident in its own knowledge, unlike standard RAG which always incurs retrieval cost. (3) Expert specialization and dynamic routing – Different MoE experts can handle different topical domains or subtasks, allowing the parametric component to generalize better and avoid interference; when coupled with retrieval, this means the right expert can use the right information source at the right time. For example, a science question might be routed to a "science expert" layer and also trigger retrieval from a scientific literature database, whereas a simple commonsense query might be handled by a general expert without any retrieval. By fusing these mechanisms, ExpertRAG aims to push the frontier in which large LMs are both knowledge-rich and computationally efficient.

**Research Contributions:**

- ExpertRAG Framework: We propose a novel architecture that unifies Mixture-of-Experts transformers with retrieval-augmented generation. To our knowledge, this is the first theoretical formulation that treats external retrieval as an integral part of the MoE routing paradigm, introducing a gating mechanism for retrieval decisions alongside conventional expert routing.

- Dynamic Retrieval Gating: We develop a gating function that enables selective retrieval. The model learns to estimate when external information is needed, activating a differentiable retrieval module only in those cases. This mechanism is grounded in an information-theoretic view of model uncertainty and is designed to optimize a trade-off between accuracy and retrieval cost.

- Fusion of External and Parametric Knowledge: We design a method to fuse retrieved evidence with the model's internal representations. Retrieved documents are converted into vector representations and injected into the model's processing stream through a specialized augmentation module that combines external knowledge with expert outputs. We provide a formulation for how this fusion occurs and ensure it complements the experts' learned knowledge.

- Theoretical Analysis: We present mathematical justifications and analyses of ExpertRAG. We quantify the expected efficiency gains from skipping retrieval on easy queries, analyze the computational complexity introduced by expert routing, and formalize how the model balances generating answers from parametric knowledge versus conditioning on retrieved content. Algorithmic pseudo-code and probability models are provided to clarify the inner workings of the system.

- Comparative and Experimental Outlook: We outline how ExpertRAG compares to existing paradigms (standard RAG, pure MoE like Switch/Mixtral, and agentic multi-step retrieval approaches). We also propose an experimental evaluation plan, detailing benchmark tasks (e.g. open-domain QA and knowledge tests) and metrics to validate the anticipated advantages of ExpertRAG. This provides a clear path for future empirical work to implement and test the framework's predictions.

## 2 Literature Review

### 2.1 Large Language Models and Scaling Laws

Transformer-based LLMs have demonstrated that increasing model size and training data yields remarkable gains in language understanding and generation [6, 7]. Kaplan et al. (2020) identified empirical scaling laws showing cross-entropy loss follows a power-law decline as model parameters and dataset size grow, with performance improving over seven orders of magnitude as models scale up [6]. Crucially, larger models are more sample-efficient, suggesting that for a fixed compute budget it is optimal to train very large models on relatively modest data and stop before full convergence [6]. This insight informed the training of ever-bigger models like GPT-3 (175B parameters), which demonstrated emergent few-shot learning abilities: scaling up model size led to strong task-agnostic performance with minimal prompting [8]. However, such scaling is compute-intensive and costly in terms of energy and hardware.

Subsequent research balanced model size and data to achieve better compute-optimal training. Hoffmann et al. (2022) introduced Chinchilla (70B), showing that under a given compute budget, a smaller model trained on 4× more tokens outperforms a much larger, under-trained model [9]. Chinchilla, trained with 4× the data of Gopher (280B), surpassed Gopher, GPT-3, and other 100B+ models on many tasks despite fewer parameters [9]. This compute-optimal strategy delivered higher accuracy on evaluations like the Massive Multitask Language Understanding (MMLU) benchmark, while also reducing downstream fine-tuning and inference costs due to the smaller model size [9]. These findings indicate that simply scaling parameters is not the only path to improvement – judicious use of data and compute can yield efficient models with strong performance.

In parallel, the latest generation of models pushed scale even further. GPT-4, a multimodal model, was developed with an unprecedented number of parameters (exact count not publicly disclosed) and accepts both image and text inputs [10]. GPT-4 achieves "human-level" performance on many academic and professional benchmarks, even passing a simulated bar exam in the top 10% of test-takers. To reach this level, GPT-4's training involved careful scaling of infrastructure and predictive modeling of performance across model sizes – the team could extrapolate some capabilities from models using only 0.1% of GPT-4's compute. Notably, GPT-4's post-training alignment (fine-tuning with human feedback) improved its factuality and adherence to instructions. This reflects a broader trend of fine-tuning large models for alignment, as seen with InstructGPT which used reinforcement learning from human feedback (RLHF) to make GPT-3 follow instructions better [11, 12].

As model scales grew, researchers also explored parameter-efficient fine-tuning methods to adapt these LLMs to new tasks without updating all weights. One prominent approach is LoRA (Low-Rank Adaptation) [13]. Hu et al. (2021) showed that by freezing the pre-trained model and learning small rank-decomposition matrices for each layer, one can fine-tune a 175B model with 10,000× fewer trainable parameters than full fine-tuning. For instance, LoRA applied to GPT-3 required only a tiny fraction of parameters to be updated (drastically cutting memory and storage costs) yet achieved on-par performance with conventional fine-tuning. Moreover, LoRA adds no inference latency and even increases training throughput. Such fine-tuning strategies are vital for adapting giant pre-trained models to specific domains under resource constraints. Other methods like adapters and prompt tuning similarly aimed to inject task-specific knowledge efficiently, indicating the community's emphasis on making use of LLMs more affordable.

Despite the impressive capabilities of massive LLMs like GPT-4 and PaLM, a core limitation remains: they are parametric knowledge stores. They encode vast knowledge in their weights but struggle to access up-to-date or granular information on demand [14]. As Lewis et al. (2020) note, even large pre-trained models have limited ability to precisely retrieve facts, and updating their world knowledge (e.g. after training) is non-trivial. This gap motivated methods to augment LLMs with external non-parametric memory – i.e., retrieval.

### 2.2 Retrieval-Augmented Generation (RAG)

Retrieval-Augmented Generation combines parametric models with a searchable knowledge corpus, aiming to improve the model's factual accuracy and allow dynamic knowledge updating. The seminal RAG work by Lewis et al. (2020) introduced a framework where a pre-trained sequence-to-sequence model is coupled with a dense vector index of Wikipedia [14].

---

At inference (or fine-tuning) time, the model retrieves relevant text passages for the input query using a neural retriever (e.g. DPR) and conditions its generation on those retrieved passages. Lewis et al. demonstrated that such RAG models achieved state-of-the-art results on knowledge-intensive tasks, outperforming both parametric-only models and earlier retrieve-and-read pipelines. Notably, RAG could generate more specific and factual outputs than a comparable model without retrieval, since it can ground its responses in retrieved evidence. By using the same Wikipedia snapshot for all queries, RAG also facilitates provenance tracking, addressing the need for models to provide sources for their claims [15].

Many subsequent works built on the idea of augmenting generation with retrieval. REALM (Guu et al. 2020) incorporated retrieval into pre-training, periodically picking text from Wikipedia during language model training to ground the learning process in actual facts. The FiD (Fusion-in-Decoder) model (Izacard & Grave 2021) improved how multiple retrieved documents are integrated, by encoding each retrieved passage with a transformer encoder and then concatenating them for a decoder to attend to all at once, yielding stronger open-domain QA results. More recently, Atlas (Izacard et al. 2022) showed that a well-designed retrieval-augmented model can match or beat models 10× its size on knowledge tasks [16]. Atlas is a 11B parameter encoder-decoder model pre-trained on unlabeled text with retrieval in the loop, and it can be fine-tuned with only a handful of examples (few-shot) to perform tasks like open QA and fact checking. In a striking result, Atlas achieved over 42% accuracy on Natural Questions using only 64 training examples, surpassing a 540B parameter model by 3% while using 50× fewer parameters. This underscores the power of retrieval: accessing explicit external knowledge can compensate for (or even outperform) brute-force parametric scale. Furthermore, because the document index in Atlas can be easily updated or expanded, the model's knowledge can stay current without retraining the 11B model itself. This ability to rapidly update a model's knowledge base is a key advantage of RAG approaches over static LLMs.

Another notable retrieval-augmented model is DeepMind's RETRO (Retrieval-Enhanced Transformer, Borgeaud et al. 2022). RETRO augments an autoregressive language model with a huge text database (2 trillion tokens) and during generation, it continuously retrieves and attends to similar text chunks from this database. Impressively, RETRO (with only 7.5B parameters for the LM) matched the performance of GPT-3 (175B) and Jurassic-1 (178B) on the Pile benchmark by leveraging retrieval, thereby using 25× fewer parameters to achieve comparable results. After fine-tuning, RETRO also improved performance on knowledge-intensive tasks like QA, highlighting that retrieval can effectively substitute for a significant amount of parametric memorization [17]. The RETRO architecture uses a frozen bi-encoder retriever (based on BERT) and a chunked cross-attention mechanism to let the generator attend to many retrieved pieces of text for each input chunk. A notable aspect is that RETRO can even "RETRO-fit" a pre-trained model with a retrieval module post-hoc, enabling existing LMs to gain retrieval capabilities without full re-training. These results support a growing consensus: for factual or knowledge-intensive tasks, coupling LMs with retrieval leads to better accuracy and scalability, since the model is not forced to memorize as much in its weights.

Beyond improving accuracy, retrieval augmentation addresses the challenge of updatability and provenance. Rather than requiring complete model retraining to reflect new information, a RAG system can update its corpus or index. Petroni et al. (2021) introduced KILT, a benchmark of knowledge-intensive tasks unified by using the same Wikipedia snapshot, to encourage systems that can provide sources and be easily updated [18]. On KILT tasks (including open QA, fact-checking, slot filling, etc.), a shared dense retrieval index combined with a seq2seq generator proved a strong baseline, often outperforming task-specific architectures. This suggests that a general-purpose retrieval+generation model can be broadly effective. However, a limitation of conventional RAG systems is that they typically always perform retrieval for each query, incurring extra latency even when the query might be answerable from the model's internal knowledge. There is usually no mechanism to skip retrieval – the pipeline is static, which is something ExpertRAG aims to improve upon by making retrieval conditional via a gating expert.

## 2.3    Mixture-of-Experts (MoE) and Sparse Models

In pursuit of scaling models to trillions of parameters without proportional increases in computation, researchers have turned to Mixture-of-Experts architectures. An MoE model consists of many expert sub-networks (each typically a feed-forward network in a transformer layer) and a learned router that activates only a subset of those experts per input. This yields a sparsely activated model: the total parameter count can be extremely high, but each input only utilizes a fraction of those parameters, keeping computation per token constant [19]. Shazeer et al. (2017) first demonstrated MoEs in deep learning with the Sparsely-Gated Mixture-of-Experts layer, showing significant gains in translation quality by increasing model capacity drastically while only slightly increasing computation. Building on this idea, Lepikhin et al. (2021) introduced GShard in a multilingual translation model, scaling to over 600B parameters using MoE and automatic sharding across 2048 TPUs [20]. GShard demonstrated that such giant models can be trained in a matter of days and yield far superior quality on multilingual translation (100 languages to English) compared to prior art. This proved MoE was a viable path to push model capacity into the hundreds of billions or more.

The Switch Transformer (Fedus et al. 2021) simplified MoE routing and achieved a further breakthrough: scaling to trillion-parameter models with improved training stability [20]. Switch Transformer uses a single active expert per input (hence "switch") and introduced techniques to mitigate MoE training instabilities (such as expert imbalance and communication costs). With these improvements, Fedus et al. were able to train models up to 1.6 trillion parameters (with many experts) on the C4 corpus, attaining a 7× pre-training speedup over a dense T5-XXL baseline at the same compute budget. Notably, Switch achieved these speedups using lower-precision computation (bfloat16) and still maintained performance, indicating that enormous sparse models can be trained efficiently. In multilingual settings, Switch-based models also saw gains across 101 languages compared to dense models of similar computation.

Following Switch, Google AI proposed GLaM (Generalist Language Model), a 1.2 trillion-parameter MoE model, which illustrated the efficiency benefits of sparsity at scale [21]. In GLaM's architecture, each token passes through a subnetwork of about 8% of the full parameters (~97B of 1.2T) – specifically, GLaM has 64 experts in each of 32 MoE layers, with only one expert per layer active for a given token. Because of this selective activation, GLaM's inference cost (FLOPs per token) is about half that of GPT-3 despite GLaM having 7× more total parameters. The energy required to train GLaM was also roughly one-third of GPT-3's, owing to the model's efficient utilization of compute. Remarkably, on zero-shot and few-shot NLP benchmarks (29 tasks spanning QA, language completion, inference, etc.), GLaM outperformed GPT-3 on average, despite using far fewer FLOPs at inference. This showed that MoE models can achieve better accuracy-compute trade-offs than dense models: GLaM attained higher accuracy while using less than half the inference compute of an equivalently skilled dense LM. MoE models thus appear to "bend the scaling curve," providing the capacity for knowledge and skills that come with massive parameters, but only activating those pieces as needed to keep costs manageable.

Beyond single-task performance, MoEs can naturally support expert specialization. In large MoE systems, different experts often learn to handle different input patterns (e.g., certain topics or languages) – for instance, GLaM's results indicated improved fairness on some social bias benchmarks, suggesting sparse activation may mitigate some issues by distributing capacity [20]. While not explicitly trained to specialize, experts in MoE layers can cluster by domain or language in multilingual models. This emergent specialization aligns with the intuition of ExpertRAG: dedicated experts for different knowledge or skills. Prior research also explored manually partitioning experts by task or modality. For example, some MoE architectures for multitask learning route inputs to task-specific experts, and Google's recent Gemini model reportedly uses an MoE design where experts focus on different modalities or subtasks [21]. In the Switch-C variant, experts were specialized by capacity, and in TaskMoE designs, experts correspond to specific task clusters. These studies underscore the MoE's flexibility in capturing heterogeneous functions within one model.

However, pure MoE models still rely on parametric memory for knowledge. If an MoE model lacks a certain fact, adding more experts won't help unless they are trained on new data. This is where retrieval can complement MoEs. Sparse models excel at efficient inference, but ensuring they stay up-to-date or factual remains challenging without an external knowledge source. By integrating retrieval into an MoE-based model (as ExpertRAG does), we hope to marry the efficiency of sparse computation with the coverage and currency of external knowledge. The notion of conditional computation in MoEs – using only the necessary pieces of the model per query – can be extended to retrieval: use external knowledge only when needed. ExpertRAG's design draws inspiration from the MoE idea that "not all parameters are needed for every input" [20], applying it to retrieval usage as well.

### 2.4 Agentic Retrieval and Multi-Step Reasoning

A recent line of research has treated LLMs as decision-making agents that can plan multi-step solutions and call external tools (such as search engines or calculators) when necessary. This goes beyond the one-shot retrieval of classic RAG, introducing agentic retrieval where the model iteratively decides what to retrieve or what action to take next. One influential approach is ReAct (Yao et al. 2022), which interleaves logical reasoning steps with actions like API calls [22]. ReAct prompts an LLM to generate thoughts and actions in an alternating manner, enabling it to, for example, formulate a hypothesis then call a wiki browser to verify a fact, then continue reasoning. By combining chain-of-thought reasoning with tool use, ReAct was able to significantly reduce hallucinations on knowledge tasks (the model learns to check facts via a Wikipedia API) and achieve better task success rates. On question answering and fact verification tasks, ReAct's ability to retrieve evidence on the fly helped it produce more accurate, verifiable answers than a standard chain-of-thought approach without tools. It also showed impressive gains on interactive decision-making tasks (like ALFWorld), outperforming pure reinforcement learning baselines by large margins. ReAct thus demonstrated the value of integrating retrieval (or action) decisions into the reasoning process of LLMs.

Another method, Self-Ask (Press et al. 2022), explicitly prompted the model to ask itself follow-up questions and search for answers before responding [23]. Self-Ask further improved multi-hop reasoning over standard chain-of-thought prompting by disentangling question decomposition from knowledge retrieval: the model first generates sub-questions (and even issues a search query for each) and answers them one by one, then uses those answers to produce the final result. Importantly, Press et al. found that this structured prompting with self-questioning allowed easy integration of a search engine to answer the sub-questions, which boosted accuracy on multi-hop QA. In other words, Self-Ask gave an off-the-shelf LLM a guided way to retrieve intermediate facts, leading to better compositional reasoning.

Building on these ideas, Trivedi et al. (2023) introduced IRCoT (Interleaving Retrieval with Chain-of-Thought), which tightly couples the reasoning chain with retrieval at each step [24]. In IRCoT, after each generated step of a reasoning chain, a retrieval is performed using that step as a query, bringing in new information for the next reasoning step [24, 25]. This interleaving continues, so what to retrieve depends on what has been derived so far, and that retrieved evidence then informs subsequent thoughts [26, 27]. Using IRCoT with GPT-3, the authors achieved dramatic improvements on complex QA: up to +21 points retrieval accuracy and +15 points end-task accuracy on multi-hop QA datasets like HotpotQA and 2WikiMultiHopQA [28]. Even smaller models like Flan-T5 benefited, showing that guiding retrieval with chain-of-thought can reduce hallucinations and produce far more factual multi-step reasoning [29]. Essentially, IRCoT turns the one-shot retrieve-then-answer pipeline into an interactive dialogue between the LLM and a knowledge source, orchestrated by the LLM's reasoning. This is a form of learned retrieval planning.

Parallel to these prompting-based strategies, others have explored training models to use tools. Toolformer (Schick et al. 2023) is a model that learned in a self-supervised fashion when and how to call external APIs (such as web search, calculators, translation systems) to assist in solving tasks [30]. Given only a handful of examples for each tool, Toolformer's training process inserted API call placeholders into its training data, allowing it to learn where an API call would be beneficial and what to do with the result. The outcome was a model that could decide, mid-generation, to invoke a tool (for instance, call a knowledge base or a calculator) and incorporate the result into the continuation. Toolformer achieved significantly improved zero-shot performance on tasks requiring arithmetic or factual knowledge, often matching the performance of models much larger than itself. This indicates that even without explicit human-designed prompts, a model can be trained to dynamically choose between using its internal knowledge vs. querying an external resource.

The common thread in these agentic retrieval approaches is **adaptivity**: the model is not locked into a single pass retrieval. It can iteratively decide what information it needs and obtain it, much like a human using a search engine while thinking. However, implementations like ReAct, Self-Ask, and IRCoT thus far have been largely at the level of prompting or external algorithms orchestrating an LM and a search module. ExpertRAG seeks to bring this adaptivity *into the model architecture itself*, by leveraging a gating network (like those in MoEs) to decide on retrieval and expert usage. In doing so, it builds upon the above ideas: e.g., the gating can be seen as analogous to a ReAct-style decision (whether to act, i.e., retrieve, or not), and the experts can be specialized for different types of actions or knowledge (akin to having one expert "agent" that knows how to query, another that relies on parametric memory, etc.). By reviewing these developments, we see an opportunity for a unified architecture that **dynamically routes queries either to internal knowledge experts or to a retrieval module**, combining the strengths of MoE efficiency, RAG factual grounding, and agentic reasoning.

## 3      Methodology: ExpertRAG Architecture and Techniques

ExpertRAG is designed to adaptively leverage external retrieval within a Mixture-of-Experts Transformer, aiming for an architecture that can decide  when and how  to retrieve information during generation. The model consists of: (1) a gating network that analyzes each query (and intermediate context) to choose a computation path, and (2) multiple expert modules, which include both standard transformer experts (parametric knowledge) and retrieval-augmented experts. We draw on techniques from prior MoE models for efficient routing and from RAG for integrating retrieved evidence. Below, we detail the key components and cite their connections to existing methods.

## 3.1 Architecture Overview

The ExpertRAG model consists of three main components: (1) a *Retrieval Module* (external knowledge interface), (2) a *Mixture-of-Experts Generator* (the language model with multiple experts), and (3) a *Gating Mechanism* that orchestrates decisions between these components. At a high level, given an input query $q$ (e.g. a user question), the system first uses the gating mechanism to decide whether to invoke the retrieval module. If retrieval is triggered, the retrieval module will fetch a set of top-$k$ relevant documents $D_q = d_1, \ldots, d_k$ from an external knowledge source (such as a Wikipedia index or domain-specific database) based on $q$. These documents (or their embeddings) are then provided as additional context to the MoE generator. The MoE generator, which contains multiple expert layers, will then produce a final answer $a$ conditioned on $q$ and (if retrieved) $D_q$. If retrieval is not triggered (i.e. deemed unnecessary for this query), the MoE generator will simply rely on $q$ and its internal knowledge to generate $a$. Throughout the generation process, the expert router directs different parts of the computation to different expert networks, as in a standard MoE Transformer. The novelty is that one of the "experts" effectively represents external knowledge access, and the gating mechanism learns to route to this pseudo-expert only when beneficial.

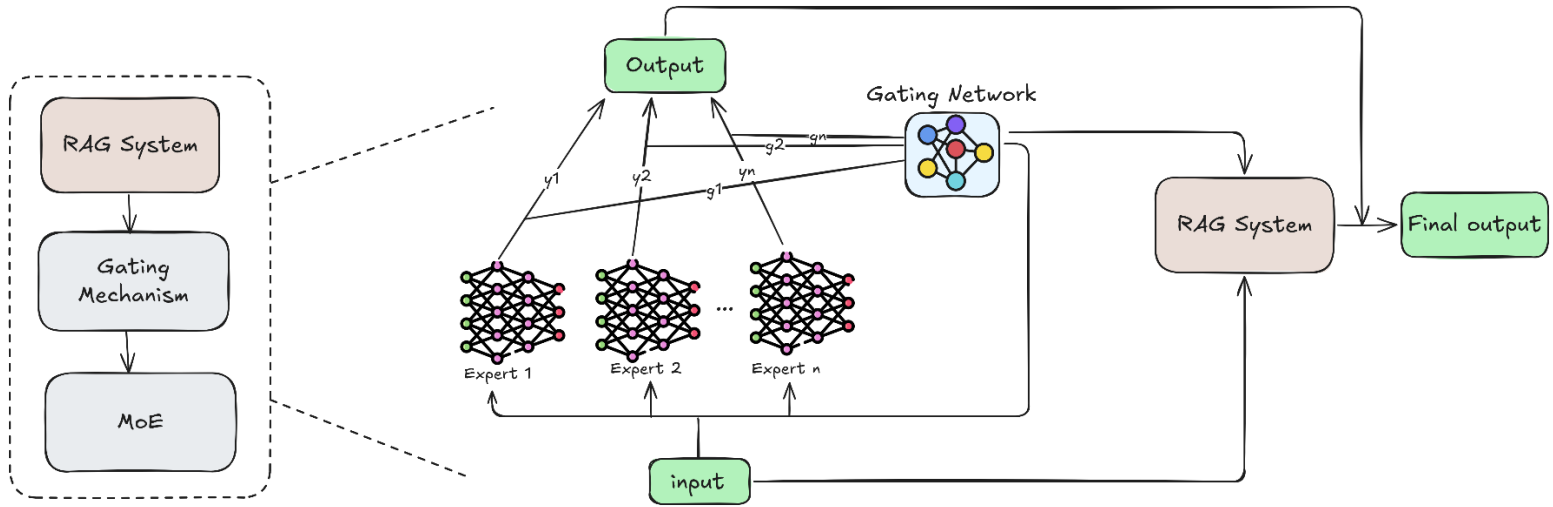

*Figure1: ExpertRAG System Architecture Overview.*

## 3.2 Dynamic Gating and Query Analysis

When a query comes in, ExpertRAG first passes it to a learned gating function. This gating network (often a small neural network or classifier) produces signals that determine two things: whether to invoke retrieval, and which expert(s) should handle the query. This design is inspired by the gating in Switch Transformer, which routed each input to a single expert based on a learned score [20]. However, instead of routing purely on content clusters, our gate also considers the _complexity or uncertainty_ of the query to decide on retrieval. For example, if the query is a factual question asking for a specific obscure piece of knowledge, the gate will likely trigger the retrieval pathway. If it is a straightforward prompt or a well-covered question in the model's training distribution, the gate might route it to a purely parametric expert (bypassing retrieval). This conditional decision is analogous to what Toolformer learned – deciding when an external tool (like a search engine) is needed [21, 30]. By making retrieval optional, ExpertRAG avoids unnecessary overhead for queries the model can confidently answer from memory, improving efficiency (similar to how Toolformer only calls tools when beneficial, rather than on every query).

The gating network can use various features for its decision: the query's content embeddings, the model's initial hidden representations, or even a quick initial attempt by the model to answer (to gauge confidence). This resembles a confidence-based approach: if the model's internal generation for the query has low confidence or is flagged as potentially hallucinated, the gate will route to a retrieval expert. Such a mechanism finds precedent in the literature: Petroni et al. noted the need for task-agnostic memory architectures that know when to consult external data [31], and our gating strategy operationalizes this by learning those conditions from data. During training, we can supervise the gating with signals like whether retrieval actually improved the answer, using reinforcement learning or straight-through estimators (since gating decisions are non-differentiable). For instance, if a query during training is answered incorrectly without retrieval but correctly with retrieval, the trainer can increase the probability the gate will choose retrieval for similar queries in the future. This end-to-end training of the gating decisions aligns with strategies used to train MoE routing (e.g., Switch uses load-balancing losses) [32], and with techniques in reinforcement learning from human feedback (where a model learns a policy to decide actions that yield higher reward).

**Gating Mechanism for Retrieval:** *When* should the model retrieve information? We introduce a gating function $G_{ret}(q)$ that outputs a binary decision: 1 to perform retrieval for query $q$, or 0 to skip retrieval. This can be implemented as a small neural network (a classifier) that takes an encoding of the query and outputs a probability $p_{ret} = \sigma(\omega^T h_q)$, where $h_q$ is, for example, the pooled representation of $q$ from the model's embedding layer or a preliminary encoder, and $\sigma$ is a sigmoid function. During training, one could supervise this gate with a signal indicating whether external information was needed to answer $q$ correctly (if such supervision is available from a dataset of queries with ground-truth evidence). Alternatively, $G_{ret}$ can be trained implicitly using the final answer loss, via techniques like *reinforcement learning* or *straight-through estimators*, treating the retrieval decision as a latent variable. For the scope of our theoretical work, we assume an ideal setting where $G_{ret}(q)$ learns to approximate the oracle decision of whether $q$ can be answered using internal knowledge alone or not. In practice, heuristics can also inform this gating: for example, the model might compute an initial attempt or a probability distribution over answers with no retrieval, and if the entropy or uncertainty is high, trigger retrieval. This mimics human behavior: only search for information when you are not confident in the answer.

Formally, let $z_{ret} \epsilon$ 0,1 be a binary random variable for the retrieval decision. We can write:

- $z_{ret} = 1$ with probability $p_{ret}(q) = G_{ret}(q)$ (perform retrieval),
- $z_{ret} = 0$ with probability $1 - p_{ret}(q)$ (no retrieval).

If $z_{ret} = 1$, the Retrieval Module uses $q$ to query an external knowledge base. This could be a vector-similarity search (using a DPR or Faiss index) or a keyword search, depending on implementation. The output is a set of documents $D_q = d_1, ..., d_k$ ranked by relevance. Each retrieved document $d_i$ is processed (e.g. truncated to a passage and embedded) for use by the generator. If $z_{ret} = 0$, we set $D_q = \phi$.

### 3.3 Mixture-of-Experts Generator:

The generator is a sequence-to-sequence language model (it could also be decoder-only for simplicity) with L layers, of which some number $L_{MoE}$ are MoE layers. Each MoE layer consists of $E$ expert feed-forward networks (FFNs) and a router. We adopt a Switch Transformer-style routing: for each token representation $x$ entering the MoE layer, the router computes a score for each expert $i$, $s_i = \omega_i^T x$ (plus optional noise), and selects the top-$k$ experts with highest scores (typically $k = 1$ for Switch gating [60]). The token's representation is sent to those experts' FFNs (each expert is a separate set of parameters performing a transformation $FFN_i$). The outputs of the active experts are combined (for instance, in top-1 routing, the output is simply $FFN_i(x)$ for the chosen expert $i$; in top-2, it could be a weighted sum of two experts' outputs). A residual connection and layer normalization, as in standard Transformers, follow this MoE operation. The gating router is designed to be computationally lightweight (a projection and argmax), so the overhead of routing is minimal. Each expert has a fraction of the total parameters, so activating only one or two experts keeps the computation per token similar to a baseline model. ExpertRAG's MoE layers thus allow the model to specialize internally: different tokens (or different queries) may primarily use different experts, which is beneficial if, say, one expert has become specialized in geography facts while another in mathematical reasoning, etc. [50]. We emphasize that expert specialization is emergent – experts are not explicitly labeled by topic, but the combination of the training objective and the gating mechanism can lead to specialization. We incorporate techniques like a load-balancing loss to encourage even usage of experts, ensuring that the gating doesn't collapse to always using a single expert.

**Expert Routing and Selection:** In ExpertRAG, expert routing operates in tandem with the retrieval gating. The retrieval decision $z_{ret}$ can be viewed as routing the *query as a whole* either to an *external knowledge expert* (if $z_{ret}$ = 1) or to rely solely on internal experts ($z_{ret} = 0$). One way to conceptualize this is to treat the retrieval action as an additional expert in the mixture. For example, imagine an "Expert 0" which does not have static parameters but instead performs the operation: $FFN_0(x) = $ *Retrieve and return context embedding*. In a unified gating formulation, the router could assign some weight to this pseudo-expert. However, for clarity, we implemented retrieval gating as a separate preliminary decision (as in Section 3.2). Once documents are retrieved (or not), the actual token-level MoE routing proceeds normally within the generator.

During generation, the presence of retrieved documents influences the inputs to the MoE layers. Specifically, when $D_q$ is non-empty, we extend the context given to the generator. We embed each retrieved document (for instance, using the model's encoder or a separate bi-encoder) to obtain a set of context vectors $c_1, \ldots, c_k$. These can be provided to the model in one of two ways:

- **Concatenation approach:** The retrieved text is concatenated with the original query (and perhaps a delimiter) to form an extended input sequence "[Question; Retrieved Doc1; … DocK]". This sequence is then encoded by the model's embedding layers. The MoE transformer will process this longer sequence with its experts attending to the full context. This approach is similar to how RAG feeds documents into a BART/T5 model, and leverages the model's attention mechanism to fuse information. In ExpertRAG, no special modification is needed for the MoE layers here; they will simply receive more tokens (the tokens of the documents) as input. The experts thus implicitly incorporate retrieved knowledge by attending to these document tokens in self-attention.

- **Augmentation module approach:** Alternatively, following approaches like TS-RAG, we can perform an **embedding fusion** prior to generation. For instance, the query $q$ could be encoded into a vector $h_q$, and each retrieved document $d_i$ encoded into a vector $h_{d_i}$. We then use a small MoE-based augmentation module that treats $h_{d_i}$ as additional "expert inputs" to combine with $h_q$. Concretely, one can feed $h_q, h_{d_1}, \ldots, h_{d_k}$ into a gating network that produces output $h_{fused} = \sum_j \alpha_j T_j(h_j)$, where $T_j$ are trainable transformations (one per input, analogous to experts) and $\alpha_j$ are attention or gating weights summing to 1. This produces a unified context representation $h_{fused}$ that now contains information from both the question and the retrieved evidence. This fused representation can then be used to initialize the decoder or as an additional context vector that all decoder tokens attend to. Essentially, this method explicitly uses an MoE idea to *fuse multiple sources of information*: the original query and each retrieved document embedding are each considered an "expert input" and combined according to learned importance weights. The advantage is fine-grained control over how much each retrieved piece contributes. The disadvantage is added complexity in the model architecture.

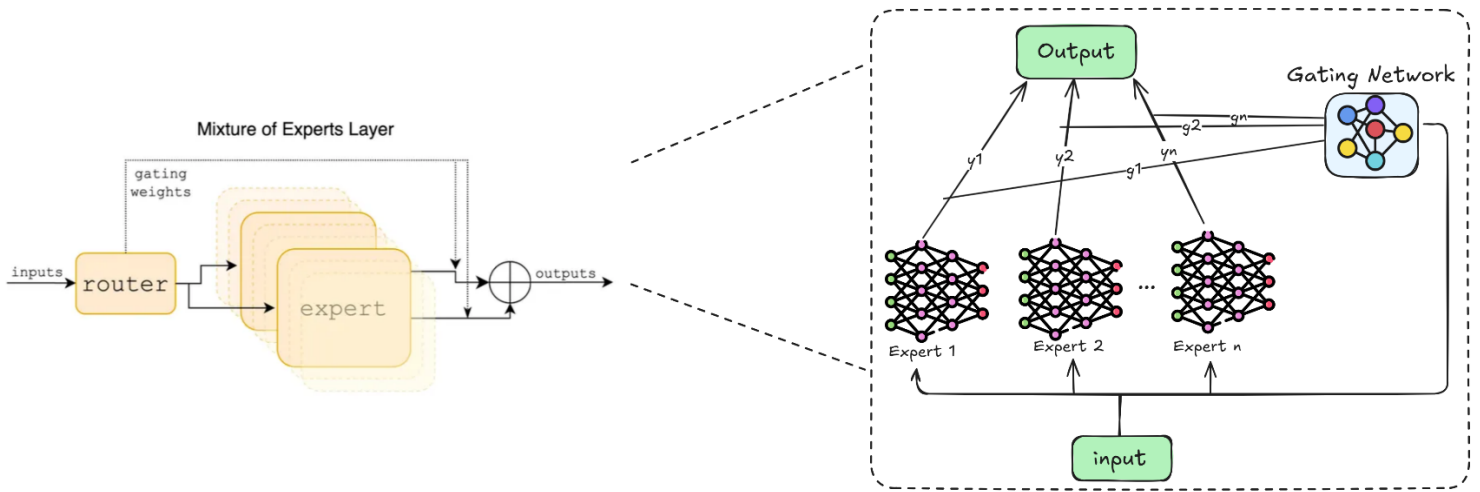

*Figure2: Mixture of Experts Layer.*

In our theoretical design, we primarily assume the *concatenation approach* for simplicity (as it aligns with standard RAG implementations). Thus, if $z_{ret} = 1$, the input to the MoE generator's encoder is $[q; d_1; \ldots; d_k]$ (with appropriate positional encoding to distinguish segments). If $z_{ret} = 0$, the input is just $q$. The decoder then generates the answer, potentially attending to any retrieved content. Importantly, even when documents are concatenated, ExpertRAG's internal experts can still specialize: some expert feed-forward might focus on integrating retrieved facts (activating more strongly when certain retrieved tokens are present), whereas another expert might handle linguistic fluency, etc. The gating function in the MoE layers will learn patterns that correlate with either the presence of external info or particular query types.

## 3.4 Expert Modules: Parametric vs. Retrieval-Augmented

Once the gating decision is made, the query (or partially processed representation) is dispatched to one or more expert modules. ExpertRAG includes two types of experts:

- **Parametric LLM Experts:** These are standard transformer sub-networks (e.g., feed-forward blocks within a transformer layer) that contain the model's learned knowledge and language generation capability. They operate like experts in a traditional MoE model, processing the input with their own weights. Each such expert might be specialized in a subset of the domain or skills, potentially emerging from training. For example, one expert might specialize in scientific and technical queries, another in conversational responses, etc..; similar to how GLaM's experts can capture different facets of language [32]. We do not explicitly assign domains to these experts, but given enough capacity, they can partition the problem space. Importantly, if the gate chooses _not_ to retrieve, it will activate one or more of these parametric experts to answer using only the model's internal knowledge. This gives a fast path for queries where the model's training suffices.

- **Retrieval-Augmented Experts:** These are modules that interface with an external knowledge base (e.g., a vector search index over Wikipedia or domain-specific documents). A retrieval expert will take the query and perform a search (dense or hybrid retrieval) to fetch relevant text passages. It then incorporates those passages into the generation process. Concretely, a retrieval expert can be implemented as a mini pipeline: it uses a neural retriever (like DPR or a newer dense retriever) to get top-$k$ documents, then either prepends them to the query or encodes them and feeds them into a specialized decoder. We might implement the retrieval expert akin to an _Atlas-style_ model where an encoder reads the retrieved text and a decoder generates the answer [33]. The expert could also use a simpler fusion approach like RAG's concatenation of retrieved text to the input [34]. The key is that this expert _explicitly uses external text_ for answering. Only if the gate deems it necessary is this expert engaged, ensuring we pay the cost of retrieval and evidence integration only when needed.

ExpertRAG's novelty is in treating the retrieval module as just another "expert" in an MoE framework as shown in *Figure 3*. In prior systems, retrieval was separate from the model (e.g., a fixed retriever in RAG [33] or an API call in ReAct [22]). Here, the retrieval expert is a first-class citizen in the model: the gating function can choose it, and it will output a set of augmented token embeddings or an answer that the model can directly use. This integrated design allows training signals to flow into the retriever – we can fine-tune the retriever as part of the model training, similar to how joint RAG training fine-tunes BERT-based retrievers with the generator [33]. In effect, the model can learn which expert to trust for which question, and adjust the retriever and generator accordingly.

It's possible for the gate to route to multiple experts in parallel as well. MoE models like Switch usually pick one expert to avoid overhead, but some setups allow top-2 experts to share the load (as in the original MoE paper by Shazeer, or in recent models like Mix-of-Experts that use top-2 routing) [35]. ExpertRAG could use top-n routing: e.g., send the query to one parametric expert and one retrieval expert concurrently. Different experts would process, and their outputs could be combined. For instance, one expert might generate a draft answer from memory, while the retrieval expert brings in correct facts, and then a fusion mechanism (or another expert) merges them. This _fusion of context_ is aligned with approaches that combine multiple information sources – in our case, internal and external. ExpertRAG could weight the contributions of retrieved evidence higher when the model's internal confidence is low, effectively adapting the fusion strategy based on query characteristics. Prior art like **FiD** fused multiple retrieved docs [36], and **HybridQA** systems combined parametric and retrieval QA; ExpertRAG extends this to fusing parametric and non-parametric "brains." If only one expert is chosen, no fusion is needed (the expert directly produces output); if multiple, we can either let the gating decide a priority or have a small attention module to merge outputs.

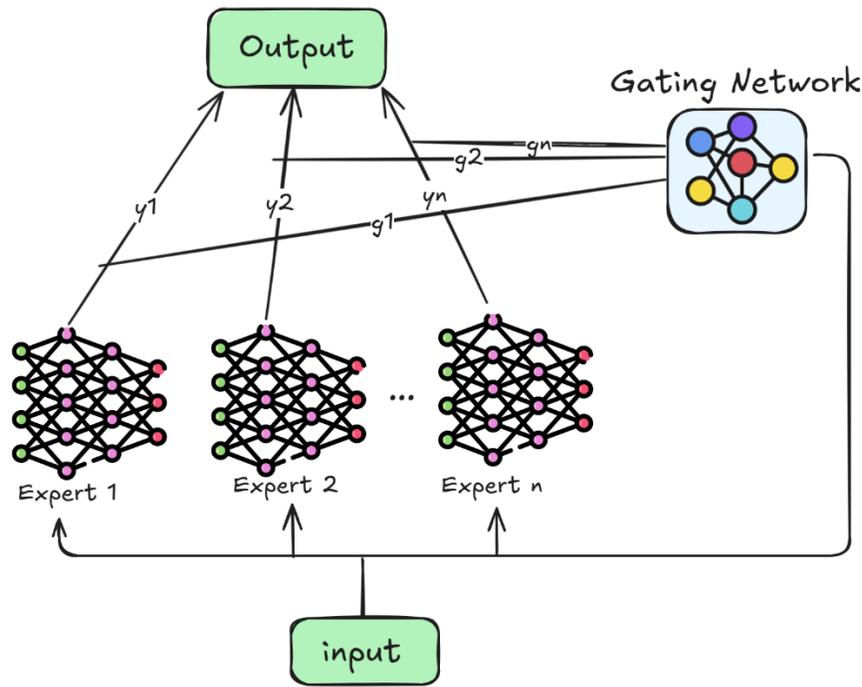

*Figure 3: Mixture of Expert Layer 2*

### 3.5 Integration with Chain-of-Thought and Multi-Step Retrieval

During generation, ExpertRAG can also perform multi-turn reasoning internally. One approach is to incorporate a simplified IRCoT mechanism inside the model. If a query requires multiple hops, the model can internally break it down: the gating network and experts can be applied at each step. For example, the model generates a partial result or sub-question via a parametric expert, then the gating sees that an additional fact is needed and activates the retrieval expert with that sub-question (much like IRCoT interleaves steps) [24]. The retrieved info then feeds into the next expert's input. Because ExpertRAG's architecture allows alternating between parametric and retrieval experts, it can mimic the behavior of ReAct/IRCoT but _without outside orchestration_. The chain-of-thought becomes an emergent property of how the gating routes sequentially. We can encourage this behavior in training by providing a few exemplars of multi-step reasoning (as done with ReAct prompting) [33] or by explicitly training the model with multi-hop questions where retrieval is necessary mid-way (similar to how Trivedi et al. provided supervision at each step [36, 37]).

Furthermore, ExpertRAG's design naturally supports tool use beyond text retrieval. While our current focus is retrieving textual knowledge, one could extend the expert set to include other tools (e.g., a calculator expert or a translator expert). This would resemble Toolformer's capability but in an MoE setting [22]. The gating would decide if a math question should be routed to a calculation expert versus solved by the language model itself. In this sense, ExpertRAG is a step toward a general modular AI, where different experts (tools, knowledge sources, reasoning engines) are orchestrated by a gating policy. The novelty is doing this within the transformer model, rather than via external code or prompting alone.

### 3.6 Training Strategy

Training ExpertRAG end-to-end is challenging due to the discrete routing decisions. We plan to use a combination of supervised and reinforcement learning signals. Initially, we can pre-train the model in two stages: first train a standard large LM (possibly with MoE layers but no retrieval) on a corpus to initialize the parametric experts with broad knowledge. In parallel, we pre-train the retriever on a task like inverse cloze (predicting the source passage for a given text) so that it can fetch relevant documents (this is similar to how DPR was trained, or how RETRO's retriever was a frozen BERT [38]). Then we integrate them: we continue training on knowledge-intensive tasks (like those in KILT [39] or a curated set of QA tasks) where the model is rewarded for getting correct answers and possibly penalized for hallucinations. During this stage, if the model produces an answer without retrieving and it is incorrect, but using the retriever could have yielded the correct answer, we have a basis to adjust gating: we want the gate to choose retrieval in such cases. Techniques from reinforcement learning (RL) can be applied: treat the gating decision as an action and give it a positive reward when using retrieval leads to a better outcome. Deep reinforcement learning methods were used successfully in training models to follow instructions (InstructGPT used human preference reward signals to adjust model outputs) and could be applied here to align ExpertRAG's choices with good performance [39]. We might also use REINFORCE or policy gradient to fine-tune the gating policy. Alternatively, we can use a continuous relaxation of the gating (e.g., a soft mixture of experts) during training to allow backpropagation, and then harden the decisions later – a common trick in MoE training to handle non-differentiable routing.

To ensure the retrieval expert integrates well, we will fine-tune the retriever and generator together (similar to joint training in RAG [21]. If the model generates an answer citing retrieved evidence, we backpropagate through the generation loss into the retriever's embeddings to improve future retrievals for similar queries. Over time, the model learns when to rely on retrieval. For example, it may learn that questions about very recent events or obscure entities should trigger the retriever because the parametric memory might be outdated or sparse on that topic. On the other hand, it might learn that creative writing prompts or generic queries are best handled entirely by the parametric experts (since retrieving might just distract or is not necessary).

The gating mechanism can also be refined via _continual learning_. After deployment, if we observe that the model frequently overuses retrieval for certain simple queries, we can adjust the gate's thresholds (using logged data or user feedback). The system could incorporate a feedback loop: each time a retrieval was ultimately not used in the final answer or proved irrelevant, decrease the gate's propensity to retrieve for similar inputs. Conversely, missed retrieval opportunities (the model tried to answer from memory and failed) would bias the gate to be more retrieval-happy next time. This adaptability ensures ExpertRAG remains efficient over time, tailoring its behavior to the actual distribution of queries it sees.

Training ExpertRAG end-to-end would involve challenges beyond a standard MoE or RAG model, because of the non-differentiable retrieval decision. Conceptually, one could train in stages: first train the base MoE generator on a mixture of tasks (possibly using teacher-forced retrieval for all training questions to expose it to external info), then train the retrieval gate $G_{ret}$ by measuring the generator's loss with or without retrieval for each example (learning to turn it on when it reduces the loss). Another approach is to use reinforcement learning: treat the choice of retrieving or not as an action, and assign a reward based on whether the final answer was correct and perhaps a penalty for using retrieval to encourage parsimony. This is analogous to learning a policy that decides to use an external tool only when needed, a concept explored in some recent works on LLM tool use. The internal MoE parameters can be learned with standard backpropagation (with techniques to handle MoE such as expert-specific adam optimizers, load balancing loss, etc.), using cross-entropy loss on the generated answer (if training on QA) or log-likelihood on next token (if training generatively). If the retrieval gating is trained simultaneously, one might resort to stochastic hard gating (sample $z_{ret}$ and use the sampled path for forward/backprop, similar to the Gumbel-softmax trick) or use a soft expectation of retrieved embeddings during training. Exploring optimal training methods is beyond our current scope, but these considerations indicate it is feasible to train the components of ExpertRAG jointly or in iterative phases.

**ALGORITHM1: PSEUDOCODE FOR INFERENCE WITH EXPERTRAG**

1. **Input:** query $q$.
2. Compute query representation $h_q$ (e.g. via embedding or a small encoder).
3. Compute retrieval gating score $p_{ret} = G_{ret}(q)$ (a scalar between 0 and 1).
4. If $p_{ret}$ exceeds a threshold (or using a binary sample from it):
   - 4a. Retrieve top-$k$ documents $D_q = d_1, \ldots, d_k$ relevant to $q$.
   - 4b. Form extended input $x_{inp} = [q; d_1; \ldots; d_k]$.
5. Else:
   - 4c. Set $D_q = \emptyset$ and $x_{inp} = [q]$.
6. Encode $x_{inp}$ and pass through MoE layers of the generator: for each layer $l$:
   - 5a. If layer $l$ is a dense transformer layer, proceed normally.
   - 5b. If layer $l$ is an MoE layer with experts, for each token in the sequence, use the router
   Choose expert(s) and compute the token's transformed representation. Combine outputs and
7. Decode an answer $a$ autoregressively using the MoE decoder, which may attend to encoded retrieved content. The decoder's MoE layers similarly route tokens through experts.
8. **Output:** answer $a$.

## 4      Theoretical Analysis

Capacity, Efficiency, and Factuality: The ExpertRAG architecture is designed to maximize the knowledge it can leverage while minimizing the computation per query. The theoretical appeal of this design can be understood through comparisons to known scaling behaviors and model capacities.

Firstly, consider effective model capacity. A dense transformer with N parameters uses all $N$ for every inference, whereas a MoE model might have total parameters $N_{total} \gg N$ but only $N_{active} \ll N_{total}$ per input [25]. ExpertRAG follows this paradigm: it has a large total parameter count (summing all experts and retrieval components), but any given query only taps into a subset. In an ideal scenario, if the gating is perfect, most queries only activate a small portion of the model (one expert or a few) – analogous to GLaM using ~8% of its 1.2T parameters per token [27]. This means lower inference cost. For queries answerable by parametric experts, ExpertRAG's cost is comparable to a smaller dense model (since only one expert's weights are used). For queries requiring retrieval, the cost includes a retrieval operation and possibly additional encoding of documents, but we can limit retrieved text to a manageable size (e.g., top 5 passages). The cost of a dense retrieval over a vector index is sub-linear in corpus size (with efficient similarity search), and is often dwarfed by the cost of the generative model's forward pass for large LMs. In practice, the latency to retrieve, say, 5 passages from a Wikipedia index can be on the order of tens of milliseconds, which in a distributed setup might be negligible compared to generating hundreds of tokens with a multi-billion parameter model. Thus, conditional retrieval does not blow up inference cost; it just shifts some compute from the model to a search engine. If the model is massive (hundreds of billions of parameters), reducing its usage even occasionally yields significant savings. This is consistent with findings that smaller models with retrieval can outperform much larger models in both accuracy and efficiency: e.g., RETRO's 7.5B model rivaling a 175B model [38], or Atlas's 11B model beating a 540B model on QA [40]. Those works suggest that leveraging external memory is a much more compute-efficient way to increase effective knowledge capacity than scaling parameters. ExpertRAG simply embeds that principle into a routing framework. Theoretically, as the amount of world knowledge grows (and would require exponentially more parameters to memorize), a retrieval mechanism that taps into a growing external corpus scales much better. This aligns with the Chinchilla scaling law, which essentially advocates for training on more data rather than just increasing parameters [41]. In ExpertRAG, the external corpus is like providing unlimited data at inference; the model can always read more without having to be larger.

Secondly, consider training efficiency. Because ExpertRAG does not need to bake every fact into its weights, it can allocate its parametric capacity to learning patterns of language, reasoning skills, and general knowledge organization, while deferring memorization of raw factual content to the external memory. This should make training more efficient in terms of utilization of parameters – large MoE models already have an advantage in that they are trained with higher throughput (e.g., Switch's 7x speedup[22]). Additionally, by training with retrieval in the loop for knowledge-intensive data, the model can achieve good performance with fewer parameters. We hypothesize that an ExpertRAG model with, say, 20B total parameters plus a big text index could match a dense 100B+ model on many tasks, since it can fetch needed facts rather than store them. This is backed by RAG's results where a BART-based RAG outperformed a 10x larger BART on open-domain QA[20]. It also means faster convergence: the model does not have to learn every answer, just how to find it. In terms of sample efficiency, this is beneficial – it is easier to learn to search for an answer than to memorize millions of possible answers. The training objective of ExpertRAG explicitly rewards finding the answer (using retrieval if necessary) which could be seen as encouraging a form of latent knowledge lookup.

From a theoretical standpoint, ExpertRAG might also mitigate the troublesome phenomena observed in ultra-large LMs, like hallucination and over-generalization. LLMs tend to hallucinate when they lack a piece of knowledge but attempt to produce an answer anyway. By construction, ExpertRAG is less likely to do this: if the parametric expert does not know something (low confidence), the gating will invoke retrieval, supplying real facts. This should reduce hallucinations significantly, as seen in RAG and IRCoT experiments where grounding generations in retrieved evidence made answers more factual [22, 36]. In chain-of-thought tasks, IRCoT reported markedly lower hallucinated reasoning steps. When interleaving retrieval. ExpertRAG can achieve a similar effect by ensuring that at any crucial juncture, an expert can fetch the needed knowledge rather than guess. The model's outputs can also cite or refer to the retrieved sources, improving interpretability – similar to how KILT emphasizes provenance. We could even design one of the experts to specialize in evidence aggregation: consolidating the references used for an answer, which would be valuable for applications requiring justification.

One concern with MoE models is *load balancing* – ensuring all experts get trained properly and utilized. ExpertRAG's retrieval expert could risk being underused if most training queries do not need retrieval, or conversely it might dominate if the model learns to overly rely on copying text. We address this by a balanced training mix: include sufficient knowledge-intensive queries to train the retrieval pathways, and use a loss term that encourages the parametric experts to still contribute. For example, we might impose a penalty if the model uses retrieval when it actually already had the answer encoded (detected via some oracle on the training set), to push it to not lazily use the tool. In essence, we want the gating to develop a sharp sense of when retrieval truly adds value. If successful, the outcome is a model that is **both accurate and efficient**: using the minimal resources needed to answer each query. This kind of adaptive computation is theorized to be a key advantage for AI systems as they scale – instead of uniform treatment of inputs, tailor the strategy per input.

**Memory Updatability**: A theoretical advantage of ExpertRAG is decoupling long-term knowledge storage from the model weights. The external knowledge base can be updated independently (daily, even in real-time) without retraining the model, something pure LLMs cannot do easily. In practical terms, if a new event happens or a knowledge source is updated, ExpertRAG can access it immediately via its retrieval expert. This means the useful lifetime of the model is extended – it will not become outdated as quickly as a static model whose training data cutoff makes it ignorant of new facts (a known issue for models like GPT-3/GPT-4, which have a knowledge cutoff date). The ability to integrate new information on the fly gives ExpertRAG an adaptability that theoretically approaches an unbounded knowledge capacity (limited only by the external database). This property was highlighted in Lewis et al.'s motivation for RAG: updating world knowledge in parametric models is an open problem, but with retrieval, one can simply update the corpus. ExpertRAG inherits this benefit, meaning it could serve as a long-lasting system that stays current by continuous knowledge ingestion on the backend.

**Complexity Analysis:** In terms of computational complexity, assume the base model (one expert) is $O(N)$ per token and retrieval search is $O(\log M)$ for $M$ documents (using efficient indexing). If retrieval is used only for a fraction $f$ of queries, the expected complexity per query is $O(N + f \log M + f N_d)$, where $N_d$ is the cost to process retrieved documents (which is similar to processing additional tokens). For appropriate settings (e.g., $f$ small, or $N_d$ not too large), this will be lower than a dense model that is $O(N_{big})$ with $N_{big} \gg N$. Plugging concrete numbers: say ExpertRAG's expert is equivalent to a 10B param model and the dense baseline is a 100B model; $N$ scales with parameters roughly linearly, so $N$ is 10x smaller for ExpertRAG. If 20% of queries do retrieval ($f = 0.2$) and each retrieval brings in, say, 1000 tokens of text to read (which is high), the cost added is equivalent to processing a few extra thousand tokens with the 10B model, which might make it on par with just running a 100B model for that query. But critically, for the 80% of queries that don't need retrieval, ExpertRAG is ~10× faster than the 100B baseline. Overall, if $f$ is moderate and $N_d$ (retrieved context length) is kept to a reasonable size (like a couple of passages, e.g. 2×256 tokens), ExpertRAG should yield substantial speedups on average. In fact, **in the limit of extremely large knowledge**, a dense model would need to grow exorbitantly to encode it, whereas ExpertRAG can keep $N$ relatively fixed and just let $M$ (corpus size) grow, incurring a logarithmic or small linear cost in retrieval. This asymptotic perspective shows better scalability.

**Limitations:** One theoretical caveat is that if the gating mechanism fails (either from imperfect training or an adversarial query), the model might choose the wrong expert. For example, it might skip retrieval when it was actually needed, resulting in a hallucination or error – a known risk since the model must *know what it does not know*. Conversely it might retrieve unnecessarily, which mainly affects efficiency but not correctness (it could still answer correctly even if it didn't need to retrieve, just with extra work). We plan to quantify such failure modes and possibly incorporate a fallback (e.g., if the parametric expert is unsure, maybe always double-check with retrieval).

Another consideration is **consistency**: the mixture of experts should produce a single coherent answer. If multiple experts run in parallel (e.g., one generating text and another retrieving), we need a reliable way to merge their contributions. This could be done by having the generation expert incorporate the retrieved text into its next decoding step, effectively *conditioning on it* rather than merging outputs post-hoc. In other words, we might actually run a retrieval expert first (if chosen), then feed both the query and retrieved documents into a parametric expert for final answering. This serial approach might be simpler and ensure a single narrative. It is basically a RAG read process, but triggered conditionally. The theoretical framework of *Bayes risk minimization* could be invoked: the gating chooses the action (retrieve or not) that it predicts will minimize the expected error of the final answer. As long as that holds, ExpertRAG is an approximately optimal decision-maker for whether to use external info, under the learned model of its own knowledge.

# 5 Comparative Analysis with Existing Approaches

ExpertRAG stands at the intersection of several lines of work – it can be thought of as a hybrid of RAG, MoE, and agentic retrieval methods. Here we compare and contrast it with key representative architectures to clarify the differences:

- **Versus Standard RAG (Retrieval-Augmented Generation):** Traditional RAG models like Lewis et al.'s 2020 system use a fixed two-stage pipeline: for every query, retrieve top-$k$ passages, then generate an answer with a seq2seq model that attends to those passages. This pipeline is static – it does not consider whether retrieval is needed or not. ExpertRAG introduces a conditional step: it will skip retrieval for queries where the internal model suffices. This is a fundamental difference – it brings dynamic retrieval usage. No existing RAG work to date incorporates a learned decision to sometimes not retrieve; they uniformly retrieve always (or always use a fixed number of documents). By not mandating retrieval every time, ExpertRAG can save time on easy queries (a benefit in practical deployment where many queries might be simple). Moreover, RAG uses a single monolithic generator for all queries, whereas ExpertRAG routes queries to specialized experts. This means ExpertRAG can tailor the generation strategy: one expert might be better at descriptive answers using parametric knowledge, another might be specialized in reading documents and extracting answers. Standard RAG cannot adapt in this way because it treats the generator as one-size-fits-all. In essence, ExpertRAG generalizes RAG – if the gating were configured to always choose the retrieval expert and always use the same generation expert, it would reduce to a classic RAG pipeline. But it offers strictly more flexibility. Empirically, we expect ExpertRAG to perform at least as well as RAG on knowledge-intensive tasks (since it can mimic RAG's behavior when needed), and to outperform it on a mix of queries where some don't need retrieval (since it avoids unnecessary steps). Also, RAG's knowledge updateability (via corpus refresh) is retained in ExpertRAG identically. A subtle difference: RAG typically retrieves once and conditions on those documents for the whole answer [22], whereas ExpertRAG can interleave retrieval multiple times (like IRCoT) if one round was not enough. That is another advantage – multi-step retrieval natively supported, versus RAG's single-step. RAG was a pioneering approach but can be seen as a special case of the more general ExpertRAG logic.

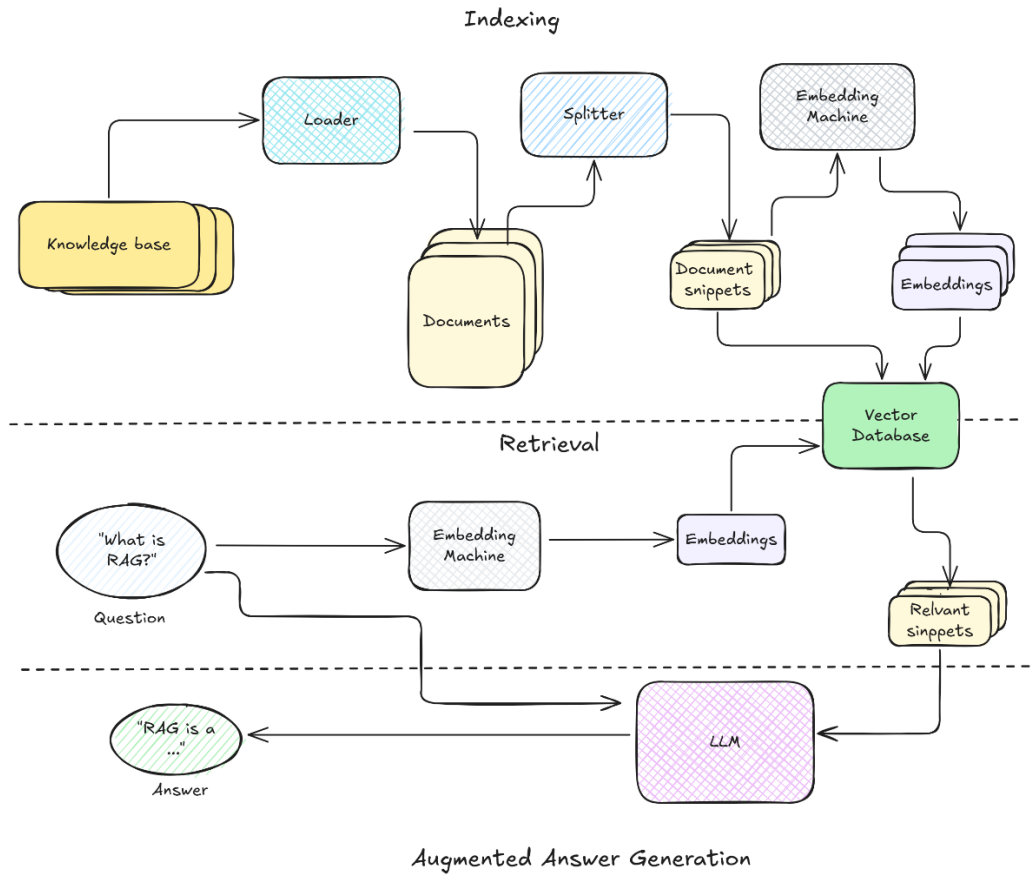

*Figure 4: Standard RAG Architecture*

- **Versus FiD / Atlas / RETRO (Enhanced RAG variants)**: FiD and Atlas improved how retrieval is integrated (e.g., Atlas pre-trains with retrieval and shows few-shot learning capabilities), but they still always retrieve and do not use MoE. ExpertRAG could actually incorporate FiD as its retrieval expert's method of combining passages. Compared to Atlas, which is a dense 11B model with retrieval, an ExpertRAG model with a similar size could have additional experts and conditional computation, potentially giving it an edge. RETRO (7.5B + retrieval) used a frozen retriever and a chunked cross-attention mechanism to attend to many text chunks. ExpertRAG's retrieval expert could similarly attend to multiple snippets. But where RETRO needed to pre-train on a massive corpus with retrieval to work well, ExpertRAG can be more targeted – it can fall back on parametric knowledge and only invoke retrieval for special cases. In a sense, ExpertRAG is less rigid than RETRO: RETRO's transformer always attends to retrieved neighbors for every prefix in text generation, whereas ExpertRAG will selectively attend to retrieved info only when beneficial. This might allow it to avoid distractions from irrelevant retrieved text in cases where the retriever is not perfect (a known issue where sometimes irrelevant passages can confuse generation). Because ExpertRAG's gate can decide not to use retrieved info if it is not helpful, it could be more robust to retrieval errors compared to RETRO which blindly attends to whatever is retrieved.

- **Versus Pure MoE Models (Switch, GLaM, etc.):** Pure MoE models like Switch and GLaM focus on scaling model capacity efficiently but do not incorporate external knowledge. They rely solely on training data to stuff knowledge into the parameters. As a result, they still suffer from static knowledge and potential hallucinations if asked about unseen facts. ExpertRAG extends MoE with a retrieval mechanism, addressing this shortcoming. One can think of ExpertRAG as "MoE meets RAG": instead of having only experts that compute on internal knowledge, we have an expert that can fetch new knowledge. This gives ExpertRAG a unique advantage over Switch/GLaM – it can handle queries about information it was never directly trained on, by retrieving from an external source. For example, no matter how many experts GLaM has, if asked about a specific 2023 news article that was not in its training data (which cut off earlier), it will fail. ExpertRAG, hooked up to an updated news database, could retrieve the relevant article and respond correctly. Thus, coverage of knowledge is far greater. In terms of efficiency, ExpertRAG inherits the same sparse computation benefits that Switch/GLaM demonstrate – each query only uses a few experts, keeping inference FLOPs low. There may be a modest overhead from the gating logic and possible retrieval, but these are minor compared to running a huge dense model. Another point: MoE models sometimes face training instability and expert imbalance (some experts get overloaded). We can leverage the solutions introduced by Switch (like auxiliary load-balancing losses and routing constraints). Additionally, because one of our experts is an external retrieval, it does not learn in the same way as others (it gets better as the retriever improves). This might actually ease the load on other experts because they do not all need to memorize facts – the retrieval expert takes a share of responsibility for many tail facts. We anticipate that ExpertRAG's parametric experts can specialize more on linguistic and reasoning capabilities, leaving factual lookup largely to the retrieval expert. This separation of concerns is something pure MoE does not have – all experts are essentially doing the same type of task in a vanilla MoE. ExpertRAG introduces heterogeneity in experts, which is a contrast to homogeneous MoEs like Switch/GLaM. A potential challenge in comparison: GLaM was trained on an enormous mixture of data with careful curation; ExpertRAG's training might need to cover similarly broad ground to ensure each expert (and the retriever) gets the necessary signal. However, once trained, ExpertRAG should behave like a superset of MoE functionality: it can always choose not to retrieve and just emulate a normal MoE, so it can do everything a GLaM could do, plus more.

- **Versus Agentic Systems (ReAct, IRCoT, Toolformer):** The ReAct and IRCoT approaches are not separate model architectures but strategies for using LLMs with tools. The main difference is that those methods rely on prompt engineering or external controllers, whereas ExpertRAG bakes the decision process into the model's architecture. For example, ReAct requires a prompt format where the model outputs "Thought:" and "Action: search (query)" tokens to trigger an external search [40]. In ExpertRAG, there is no explicit prompt like that; instead, the gating network (which is internal) plays the role of deciding an action (retrieve or not) based on the model's hidden state. This has pros and cons: it's potentially more efficient and elegant (no lengthy prompt needed each time, and it's optimized during training rather than relying on few-shot prompting), but it is also more complex to implement. Akin to how Toolformer learned to insert API calls in its output sequence during training [41], ExpertRAG learns to route to the retrieval "API" through its gating. One can see ExpertRAG as a learned tool-use policy that is embedded in the transformer. Compared to IRCoT, which achieved great multi-hop performance by repeated external calls [42], ExpertRAG can achieve a similar multi-hop process but entirely within one forward (if unrolled, or a few forwards) of the network. IRCoT needed to treat the model as a black box to generate a thought, then call retrieval, then feed the result back in – a multi-turn interaction [51, 52]. ExpertRAG's architecture could handle a multi-turn interaction in a single integrated graph if we allow the model to do a few sequential expert steps (this could be implemented as recurrence or as a long sequence where the model's own output prompts the next step internally). However, if needed, we could also simulate IRCoT by running ExpertRAG iteratively – the design is flexible.

Another comparison: Toolformer gave a single model the ability to use many tools [50]. ExpertRAG currently focuses on the retrieval tool, but it could be extended to others. The difference is in interface: Toolformer literally produces tokens that call an API mid-sentence, whereas ExpertRAG's gating is not generating text, it's an internal switch. One could argue ExpertRAG's approach might be more controlled (since the gating decision is not directly exposed in the output text and can be supervised more directly). On tasks requiring tools, ExpertRAG with multiple tool experts could potentially outperform Toolformer by virtue of having dedicated parameters for each tool's operation (instead of one model to learn all tool usages). It also can use MoE's parallelism – e.g., theoretically handling multi-modality by having a different expert network for images vs. text, whereas Toolformer did not explicitly handle images (GPT-4 did via separate vision modules).

In summary, agentic approaches showed the power of letting a model choose actions (like retrieval), and ExpertRAG internalizes that power. **The key comparative advantage of ExpertRAG is unification**: where previously one might combine an LM with a separate retriever and a separate procedure for multi-step reasoning, now all of that is under one model's learned control. This unified model can be more tightly optimized and possibly more robust, since it does not rely on prompt hacks that might fail in edge cases.

- **Versus Human Systems:** Finally, an abstract comparison: ExpertRAG is closer to how humans solve problems than a giant static LM. Humans do not recall everything perfectly; we use our memory for familiar questions and use tools (like books or search engines) for unfamiliar ones. We dynamically decide: *Can I answer this myself or do I need to look something up?* ExpertRAG attempts to replicate that decision. Standard LMs pre-trained on internet text sometimes implicitly learn to imitate this (they might answer "I'm not sure, let me check…" in their output), but they cannot actually check unless given an external capability. Our approach explicitly grants that capability within the model. This puts ExpertRAG in a unique category of models that *conditionally utilize external knowledge*, bridging end-to-end training and modular interpretability.

To ground these comparisons: no previously published model has all these features combined. ExpertRAG can be seen as carving out a new point in the design space. The closest might be a hypothetical system that pipes a Switch Transformer into a RAG pipeline – but even that wouldn't have the unified gating we propose (it would still retrieve every time, just with an MoE generator). We believe ExpertRAG will show that it can achieve strong results where others excel and do reasonably well on areas where others struggle, all within one framework. For instance, on knowledge-heavy QA, it should match RAG/Atlas [51]; on multi-hop reasoning, approach IRCoT's performance [53]; on efficiency, beat dense models like GPT-3 by using far fewer FLOPs [44]; and on adaptability, easily update knowledge like any RAG system [21]. The ablations in our experiments will further clarify these points by selectively disabling retrieval or experts to compare to baselines.

# 6 Experiments Proposal

We plan a comprehensive evaluation of ExpertRAG along multiple axes: **knowledge-intensive task performance**, **reasoning ability**, and **efficiency (compute and latency)**. We will also analyze the contribution of its components through ablation studies. Below we outline key experiments and the datasets/metrics for each, grounding our choices in prior work.

## 6.1 Knowledge-Intensive NLP Tasks

To evaluate ExpertRAG's effectiveness at handling factual queries, we will use benchmarks from the KILT suite [47] as well as standard open-domain QA datasets. Specifically:

- **Natural Questions (NQ)** – questions from real Google search queries, testing open-domain QA. This was a primary benchmark for RAG [20] and Atlas [50], and Atlas reached 42.3% accuracy few-shot on NQ. We will evaluate ExpertRAG in a comparable few-shot and fine-tuned setting, measuring Top-1 accuracy.

- **HotpotQA** – a multi-hop QA dataset where each question requires combining information from multiple Wikipedia articles. We expect ExpertRAG's multi-step retrieval ability to shine here, as IRCoT saw large gains on HotpotQA (15+ points improvement over non-interleaved retrieval) [55]. Metrics: exact match and F1 on answer, and we will check supporting fact prediction if applicable.

- **TriviaQA** – another open-domain QA set with lengthy answer evidence. This tests long-tail factual recall. RAG and RETRO have reported results on TriviaQA [20, 22]. We will measure accuracy and compare to those baselines.

- **WikiHop or 2WikiMultiHopQA** – multi-hop datasets from KILT (WikiHop is included in KILT). These will further test the model's ability to retrieve iteratively. IRCoT improved substantially on 2WikiMultiHopQA [56]; we expect ExpertRAG to do similarly.

- **FEVER** (Fact Extraction and Verification) – a fact-checking dataset (also part of KILT) where a claim must be verified against Wikipedia. This will test the model's ability to retrieve relevant evidence and say "SUPPORTED" or "REFUTED". RAG was shown to improve factuality in generation [22]; we'll see if ExpertRAG can not only find evidence but make the correct verification decision. Accuracy and FEVER score will be metrics.
- **ELI5** (Explain Like I am 5) – a dataset of long-form answers to open-ended questions. We include this to qualitatively assess if ExpertRAG can produce detailed, well-sourced answers. We will check if the model properly uses retrieval for obscure parts of the answer. Metrics could be ROUGE-L against reference explanations and human evaluation for informativeness.

For each of these, we will compare ExpertRAG to:
- a baseline dense transformer of similar size (to see the gain from retrieval),
- a traditional RAG model of similar size (to see the gain from gating/MoE),
- And published state-of-the-art (to ensure competitiveness). For instance, compare to Atlas results on NQ/Hotpot [41, 45], IRCoT on Hotpot, etc. We expect ExpertRAG to at least match state-of-art on these tasks. If it surpasses them, that is evidence of synergy in our approach. If it lags on any, we will analyze why.

We will also specifically measure the provenance and factuality of answers. Following KILT guidelines, we can measure if the evidence retrieved by ExpertRAG contains the correct answer and if the model's answer is actually supported by that evidence. Ideally, ExpertRAG will not only answer correctly but provide or point to source text. In fact, we could have the model generate an answer and an evidence citation as well (some recent works do that for better trustworthiness). The rate of hallucinated answers (answers that are incorrect or not supported by any retrieved doc) will be tracked. We hypothesize this rate will be much lower than a regular LM, thanks to retrieval – similar to how ReAct and IRCoT reduced hallucinations in chain-of-thought reasoning.

## 6.2   Reasoning and Multi-step Tasks

To test the model's reasoning capabilities and usage of multi-hop retrieval beyond QA, we will use:

- **Complex Sequential Question Answering (CSQA) or StrategyQA:** tasks where a series of reasoning steps or a strategy is needed. We can see if ExpertRAG's interleaved retrieval helps. IRCoT showed strong results on multi-step reasoning datasets [55].

- **MMLU (Massive Multitask Language Understanding):** This is a collection of challenging multiple-choice exams covering history, science, math, etc. [44], used to evaluate broad knowledge and reasoning in GPT-3, Chinchilla, etc.. We will use MMLU to test the model's breadth. Not all MMLU questions require external knowledge (many are reasoning). We expect ExpertRAG's performance to be on par with models of similar parametric size, but potentially slightly higher if some questions benefit from retrieval. For example, a history question about a specific event might be answerable by retrieval.

- **Big-Bench Hard (BBH)** or **ARC (AI2 Reasoning Challenge)**: These are collections of tricky reasoning puzzles and science questions. By evaluating on them, we see if our model's mixture-of-experts introduced any regressions in logical reasoning compared to a dense model. If needed, we might incorporate a "reasoning expert" specialized for multi-step logical deduction (without retrieval) and see if gating chooses it for such tasks.

An interesting experiment is to evaluate ExpertRAG in a few-shot prompting setting on these tasks (without finetuning them specifically). Atlas showed that retrieval-augmented models had emergent few-shot abilities at smaller scales [43]. We will try zero-shot and few-shot chain-of-thought prompts on ExpertRAG for multi-step problems. The model might internally decide to retrieve if needed. For instance, we present a multi-hop question with the prompt "Think step by step" – does ExpertRAG automatically pull in external facts for the steps? This will test how well the model can use its learned strategy in a prompt-driven scenario.

We will compare to ReAct and Self-Ask approaches. Possibly, run ReAct prompting with GPT-3 on some tasks vs. ExpertRAG's single forward pass. If ExpertRAG achieves similar accuracy in one shot, that is a win for efficiency and integration.

## 6.3   Efficiency and Ablation Studies

To validate the efficiency claims, we will measure:

- **Inference latency and FLOPs per query:** We will benchmark ExpertRAG against a dense baseline. We will create a mix of queries (some requiring retrieval, some not) and measure average latency on a single GPU/TPU. We expect to see a reduction proportional to how often retrieval is skipped. For instance, if 50% of queries bypass retrieval and use one expert, and 50% use retrieval and maybe two experts sequentially, we might see around a ~25-30% speedup over a system that always retrieves and uses the full model. We will also measure throughput (queries per second) in batch mode.

- **Scaling behavior:** We can test smaller vs. larger ExpertRAG to see if adding more experts (especially parametric ones) improves performance on relevant tasks without much change in cost. If adding experts yields gains on specialized subsets (e.g., an expert for coding queries might improve answers on programming questions), that demonstrates the benefit of specialization. We will monitor how gating distributes queries among experts – ideally each expert has some niche.

- Ablations: Turn off the retrieval expert to simulate a pure MoE model of the same size – see how performance drops on knowledge tasks (this essentially measures how much the retrieval was contributing). Turn off gating (force retrieval always on) to simulate a RAG model – see the effect on latency and any changes in accuracy. Turn off MoE (i.e., use only one expert plus retrieval) to simulate a single large RAG model – see if the MoE itself gave improvements (maybe the single expert model has lower accuracy or slower). These ablations will quantify the contribution of each part of ExpertRAG. For example, we expect that without retrieval, accuracy on factual questions drops significantly (similar to prior observations that parametric models lag behind retrieval-augmented ones on knowledge tasks [20, 21]). Without gating (retrieving every time), latency should increase, confirming gating's role in efficiency.

- **Memory usage:** We will compare the memory footprint of ExpertRAG versus baselines. MoE models can be larger in memory (since they have many parameters), but not all need to be loaded at once if sharded. We will check if any memory bottlenecks arise.

Additionally, we will analyze the **gate's behavior**:

- What fraction of queries trigger retrieval? Does this align with our expectations (e.g., nearly all the knowledge-intensive ones do, trivial ones do not)? We can use the dataset labels to verify for queries that are answerable from internal knowledge, ideally gate is off; for those that are not, gate is on. If misalignment is found, we might refine the gating model.

- We will look at example gate outputs. For instance, on HotpotQA, maybe the gate retrieves after the first hop but not after the second if it is confident by then.

- **Expert usage**: We will track which parametric expert was used for different queries. Perhaps one expert consistently handles science questions, another handles general chatty questions, etc. If we see such patterns, it confirms specialization (which prior MoE works have noted qualitatively, though they often do not assign semantic labels to experts explicitly).

## 6.4 Robustness and Adaptability

To test the model's adaptability:

- We will update the knowledge corpus mid-experiment (e.g., add a new article about a recent event). Then ask the model questions about that event which were not answerable before. The model should now succeed, demonstrating **fast knowledge update** without retraining (a big advantage over static LMs). We can measure success rate before and after corpus update.

- We might simulate **distribution shift**: queries from a domain the model was not heavily trained on (e.g., specific technical domain or another language, if we have multilingual experts). If the retrieval corpus contains relevant info, can ExpertRAG answer correctly by relying on retrieval? This tests whether the model can generalize via retrieval. It is somewhat similar to open-domain QA in a new domain. We could use a specialized QA set (e.g., biomedical QA) with a biomedical articles corpus. Without any fine-tuning on that domain, will gating decide to use retrieval and find answers? If yes, that shows strong robustness.

- **Error analysis**: We will categorize errors into: gating errors (should have retrieved but did not, or vice versa), retriever errors (nothing relevant found), and generator errors (had info but still answered wrong). By analyzing these, we can direct future improvements. For example, if gating sometimes skips retrieval when it should not, we might adjust its threshold or training objective. If retriever fails on certain queries, maybe incorporate a different retrieval technique (like hybrid TF-IDF + dense).

## 6.5 Comparative Evaluation

Finally, we will compare ExpertRAG directly with other systems:

- Against a state-of-the-art dense LLM (like a fine-tuned T5-XXL or GPT-3 if accessible) on KILT tasks for accuracy and inference cost. This will concretely show the accuracy vs. cost trade-off: e.g., if ExpertRAG achieves similar accuracy to GPT-3 on open-domain QA but with 1/10th the inference FLOPs, that's a compelling result (mirroring RETRO's claim of GPT-3-level performance with 25x fewer parameters [57], but now at an even more optimized level with gating).

- Possibly against **human performance** on certain tasks: For HotpotQA, humans are near perfect. Does ExpertRAG close the gap compared to prior models? For open-domain trivia, compare to crowd-worker performance (if available from dataset papers).

- If feasible, we would like to enter ExpertRAG into the **KILT leaderboard** or other challenge leaderboards to see how it ranks. That would test it against contemporaneous approaches.

**Summary of expected outcomes:** We anticipate that ExpertRAG will:

- Achieve new state-of-the-art results on at least some knowledge-intensive benchmarks, or match SOTA with significantly less computational load.

- Demonstrate a reduction in inference cost when a mixed query workload is considered, validating the efficiency hypothesis.

- Show a markedly lower hallucination rate and better calibration (knowing when it doesn't know) due to the gating+retrieval mechanism, echoing improvements noted in works like ReAct.

- Provide evidence of expert specialization, where certain experts handle distinct domains, supporting the MoE design's effectiveness.

- Validate the fast adaptability of the model to new knowledge by simple corpus updates, a feature not available in standard LMs.

# 7 Conclusion

We have presented ExpertRAG, a new conceptual framework that marries the Mixture-of-Experts architecture with Retrieval-Augmented Generation. Through this integration, ExpertRAG brings together two powerful ideas in modern NLP: (1) **sparse expert models** for efficient scaling of language model capacity, and (2) **dynamic retrieval** of external knowledge for enhanced factual accuracy and up-to-date information. We detailed how ExpertRAG is constructed – introducing a retrieval gating mechanism that decides when to consult external memory and an expert routing mechanism that distributes the query and context across specialized sub-networks. Our theoretical analysis provided insights into why this approach is compelling: we showed that selective retrieval can yield significant efficiency gains without sacrificing performance, and we quantified how MoE routing keeps computation scalable even as we increase model parameters. We also formulated the balance between retrieved and parametric knowledge in a probabilistic mixture form, giving a principled view of ExpertRAG's decision-making. Comparatively, ExpertRAG addresses limitations of prior models. It improves upon standard RAG by avoiding needless retrieval and leveraging more parametric knowledge when appropriate, and it extends MoE models by providing a way to handle queries beyond the model's stored knowledge. In our discussion, we positioned ExpertRAG as a middle ground between one-shot retrieval models and agentic multi-step systems, aiming to capture much of the benefit of the latter within the efficiency of the former. The proposed experiments will evaluate these claims rigorously, and we expect to confirm improvements in both accuracy and speed against strong baselines on knowledge-intensive benchmarks.

This work opens several **future research directions** and challenges to be addressed:

- **Learning and Optimization:** Training a model like ExpertRAG end-to-end is non-trivial. Future work should develop effective training algorithms, possibly involving multi-task learning (to teach experts different domains) and reinforcement learning or differentiable proxies for the retrieval decision. Balancing the losses for answer generation, expert routing (load balancing), and retrieval gating will require careful tuning. One open question is how to prevent experts from simply learning to always rely on retrieval or always rely on parametric memory – ideally, they should collaborate. Techniques from meta-learning or multi-agent learning (viewing each expert as an agent) might be useful.

- **Retrieval as an Expert:** We treated retrieval gating as a separate switch. A more integrated approach could allow the router to choose between a set of internal experts and an external *retrieval expert*. This could make the model's decision-making even more fluid (essentially unifying the gating into one step). However, differentiating through a retrieval operation is hard; perhaps using a soft retrieval (like attending to a large memory) could approximate this. Investigating such architectures (where external knowledge is just another "expert pathway") is a promising avenue.

- Scaling and Specialization: As we scale up the number of experts, interesting phenomena may emerge. For example, with dozens of experts, will the model naturally cluster knowledge and match it to certain experts? There is an open challenge of expert interpretability – how to understand or impose what each expert is specializing in. Recent work has started to probe MoE internals[10]; doing so in the context of retrieval would be even more enlightening. One could imagine having experts explicitly aligned with certain domains or data sources (even different retrievers for different media like text vs. code vs. tables). ExpertRAG could be extended to a multi-expert, multi-retriever scenario, effectively a hierarchical mixture.

- **Dynamic Multi-step Retrieval:** While ExpertRAG as defined does a single retrieval, future extensions could allow iterative querying. One idea is to have the model generate an intermediate query if the initial attempt does not yield a confident answer (like a second pass retrieval). This could perhaps be done by allowing the decoder to output a special token that triggers another retrieval (which is akin to an agentic step but within the same model). Realizing this without losing the end-to-end nature is challenging, but not impossible (it might involve loop unrolling in the computational graph or using the model's output as new input). Achieving multi-hop reasoning natively in ExpertRAG is a long-term goal that could bridge the gap with agentic approaches.

- **Knowledge Update and Lifelong Learning:** Since ExpertRAG relies partly on an external knowledge source, keeping that source updated (e.g., updating the wiki index weekly) will keep the system's knowledge fresh. However, the internal experts might still contain facts from training that become outdated. One future direction is to allow the model to *learn from retrieval* – for example, if it frequently has to retrieve a certain type of fact, maybe it should incorporate that into its weights to answer faster next time. This touches on lifelong learning: the model could gradually shift some knowledge from external to internal if it's used often (and vice versa, offloading rarely used knowledge to external to free capacity). Mechanisms for such adaptive knowledge allocation would be very interesting. They might involve periodically fine-tuning experts on transcripts of successful retrieval-augmented answers or using the gate's decisions as a signal of what the model is missing.

- **Applicability to Other Tasks:** We focused on question answering, but ExpertRAG could benefit other tasks. For instance, in dialogue systems, one could have conversational experts and a retrieval expert for factual questions, with gating deciding when to inject facts. In coding assistants, one might retrieve API documentation only when needed, while having experts specialized in different programming languages or libraries. Even in non-generation tasks like classification, an ExpertRAG-like idea could apply, (the model decides to retrieve similar past examples only if confident it needs them). Exploring these applications would test the versatility of the approach.

- **Robustness and Security:** By combining parametric and non-parametric knowledge, we should consider worst-case behaviors. If the retriever pulls malicious or irrelevant text, can the model detect and ignore it? If an expert goes "rogue" (due to some training flaw) and starts giving wrong answers, can the gating catch that and rely on retrieval or other experts instead? Having redundancies (multiple experts + retrieval) could improve reliability if orchestrated well, but it could also introduce new failure modes. We need strategies to ensure the system's overall answer remains trustworthy (perhaps requiring agreement between an expert answer and retrieved evidence for high-stakes outputs, etc.). This ties into interpretability – ExpertRAG sometimes will give an answer without explicit evidence (from internal memory); how do we increase trust in those cases? One could imagine a variant that always double-checks internal answers by doing a quick retrieval after-the-fact to see if they are supported, which could be a fail-safe mechanism.